# Temperature dependence of hydrogen local mode frequencies in the ordering bcc metal hydrides

I.G.Ratishvili [a] *, N.Z.Namoradze [b]

[a] E.Andronikashvili Institute of Physics, Tbilisi, Georgia,
[b] Institute of Cybernetics, Tbilisi, Georgia

*Corresponding author
E-mail address: ratisoso@hotmail.com

**Abstract**

A simple physical conception explaining the experimentally observed temperature dependence of the local mode frequencies in the ordering metal hydrides is proposed. Calculations are performed for *bcc* vanadium deuterides. The model can be applied to other physical phenomena which are related to the hydrogen motion in metal lattices.

**Introduction**

Local modes associated with interstitial hydrogen vibrations in the metal lattices were intensively investigated in the 70-th and 80-th years. There were established some obvious features of the local modes (such as the level splitting in the crystalline fields of non-cubic symmetry [1], or the mass-dependent reduction of the local mode frequency on passing from hydrogen to deuterium atoms [2, 3]), as well as many peculiar details of these phenomena. Particularly, on investigating the local modes of hydrogen, deuterium and tritium isotopes in the β-phases of niobium hydrides some valuable information concerning the corresponding potential wells was obtained [2]. The influence of hydrogen ordering on the local modes and on the anharmonicity parameters of local vibrations was emphasized in [3]. Extremely dilute alloys of hydrogen in V, Nb and Ta were examined in [4] and it was shown, particularly, that in $NbH_{0.0055}$ at temperature lowering from 295 K to 210 K the first (lower) local frequency changes from 106 to 118 meV, while the upper one (equal to ~ 163 meV) remains practically unchanged.

We intend to analyze more precisely the role of hydrogen ordering process on the local modes of hydrogen atoms in *bcc* metal lattices and, particularly, estimate the corresponding shift of appropriate energy levels. We shall base on the experimental data [5] obtained for ordering $V_2D$ compound at temperatures 425 K and 300 K, i.e. sufficiently above and below the order-disorder transition temperature $T_c [V_2D] = 406$ K [6]. From measurements [5] it follows that at temperature lowering both local frequencies, $\omega_1$ and $\omega_2$, or, equivalently, the corresponding Einstein temperatures $\theta_E^{(1)} = (\hbar\omega_1 / k_B)$ and $\theta_E^{(2)} = (\hbar\omega_2 / k_B)$, *become lower* in contradiction with the mentioned above results [4] obtained for dilute alloys. From [5] it follows

$$\theta_E^{(1)}(425 \text{ K}) = 936 \text{ K}, \quad \theta_E^{(2)}(425 \text{ K}) = 1408 \text{ K}, \qquad (A)$$

$$\theta_E^{(1)}(300 \text{ K}) = 561 \text{ K}, \quad \theta_E^{(2)}(300 \text{ K}) = 897 \text{ K}, \qquad (B)$$

At the first glance, it will be more physically acceptable the inverse situation. Indeed, in the ordered state hydrogen atoms are usually assumed to be located in the deeper potential wells than in the disordered state [7, 8]. Basing on this assumption it has to be expected that they should be bound to metal atoms more tightly. This effect has to be manifested by the increase of local frequencies at lowering temperatures (as it was registered, particularly, in the dilute alloy $NbH_{0.0055}$ [4]). Nevertheless, experimental results [5] establish an opposite temperature dependence. We shall try to explain this paradox on applying a specific model.

**The system under consideration**

It is well known that equilibrium phase diagrams of vanadium deuterides and vanadium hydrides are sufficiently different [9]. We consider especially the deuteride $V_2D$ which has a relatively narrow monophase region on the concentration scale. The corresponding phase diagram is presented in Fig. 1 [9]. It is established experimentally that in the case of this compound interstitial hydrogen atoms at high temperatures (in the α-phase) are distributed randomly between all tetrahedral interstitial positions, while in the ordered β-phase they are collected *in one bcc sublattice of octahedral positions* and form an ordered configuration (see the review articles [7, 9]). The disorder-order transition occurs at temperature $T = T_c = 406$ K. The ordered configuration is described by the wave vector $\mathbf{k}_1 = (2\pi / a)(0 \; \frac{1}{2} \; \frac{1}{2})$ [10, 7].

The corresponding superlattice is represented by the distribution function [10, 7]:

$$n(x, y, z) = c + \eta(T) \gamma \exp[i \mathbf{k}_1 \mathbf{R}_i] = c + \eta(T) \gamma \exp[i\pi(y + z)]. \qquad (1)$$

Here (x, y, z) denote the coordinates of sites of the *bcc* lattice formed by the positions of the occupied octa- sublattice, $n(x, y, z)$ is the *occupation number* of the given (x, y, z)-site (which characterizes the probability of occupation of the given site by hydrogen atoms, H-atom). Concentration $c$ is determined as the ratio of the number of hydrogen atoms $N_H$ to the number of sites N in the given *bcc* interstitial sublattice. In the case of an ideal metal lattice, N equals to the number of metal atoms $N_M$, and then we can write: $c = (N_H / N_M)$. $\eta(T)$ is the temperature-dependent order parameter and $\gamma$ is the normalizing factor.

On the set of *bcc* lattice sites function (1) takes two different values,

$$n(x, y, z) = c + \eta(T) \gamma \equiv n_1 \quad \text{and} \quad n(x, y, z) = c - \eta(T) \gamma \equiv n_2 . \qquad (2)$$

If in the stoichiometric alloy with $c = 0.5$ the completely ordered state is associated with the order parameter value $\eta(T) = 1$, and it is represented by the set of occupation numbers [$n_1 = 1$, $n_2 = 0$], then we have to assume additionally that the normalizing parameter $\gamma$ equals 0.5 ($\gamma = 0.5$).

The ordering subsystem is characterized by the free-energy function [10]

$$F(\eta; c, T) = E(\eta; c) - T S(\eta; c), \qquad (3a)$$

$$E(\eta; c) = (1/2) \sum_{ij} U_{HH}(\mathbf{R}_i - \mathbf{R}_j) n(\mathbf{R}_i) n(\mathbf{R}_i), \qquad i, j = 1, \ldots . N \qquad (3b)$$

$$S(\eta; c) = - k_B \sum_i [n(\mathbf{R}_i) \ln n(\mathbf{R}_i) + (1 - n(\mathbf{R}_i)) \ln (1 - n(\mathbf{R}_i))]. \quad i = 1, \ldots N \qquad (3c)$$

In equation (3b) $U_{HH}(\mathbf{R}_i - \mathbf{R}_j)$ denotes the interaction between hydrogen atoms located in the sites *i* and *j*, and $n(\mathbf{R}_i)$ and $n(\mathbf{R}_j)$ are the occupation numbers and characterize probabilities of occupation of the corresponding sites by H-atoms. Introducing expression (1) in (3b) and (3c) we obtain, respectively,

$$\sum_{ij} U_{HH}(\mathbf{R}_i - \mathbf{R}_j) n(\mathbf{R}_i) n(\mathbf{R}_i) = N [U(0) c^2 + U(\mathbf{k}_1) (\eta\gamma)^2], \qquad (4)$$

$$U(0) = (1 / N) \sum_{ij} U_{HH}(\mathbf{R}_i - \mathbf{R}_j), \qquad (5a)$$

$$U(\mathbf{k}_1) = (1 / N) \sum_{ij} U_{HH}(\mathbf{R}_i - \mathbf{R}_j) \exp [i \mathbf{k}_1 (\mathbf{R}_i - \mathbf{R}_j)], \qquad (5b)$$

$$S(\eta; c) = - N k_B \sum_i \nu_i [n_i \ln n_i + (1 - n_i) \ln (1 - n_i)]; \quad i = 1, 2 \qquad (6)$$

In (6) $\nu_1$ and $\nu_2$ denote the partial number of *bcc* lattice sites where $n(x, y, z)$ equals $n_1$ and $n_2$, respectively. In the case of distribution function (1) we have: $\nu_1 = \nu_2 = 0.5$.

It is preferable to determine the energy constants U(0) and U($\mathbf{k}_1$) in the temperature units. Below we shall use the energy parameters $(U(\mathbf{k}_1) / k_B) \equiv V(\mathbf{k}_1)$ and $(U(0) / k_B) \equiv V(0)$. Then the H-subsystem ordering energy can be presented as [11]

$$E(\eta; c) = (1/2) N k_B [V(0) c^2 + V(\mathbf{k}_1) (\eta\gamma)^2]. \qquad (7)$$

The equilibrium states of the ordering hydrogen subsystem are described by the order parameter values $\eta(T)$ which are determined as the coordinates of the free-energy minimum in the $\eta$-space at the given temperature T. In order to obtain the extremum values of order parameters we have to solve the equation $\partial F / \partial \eta = 0$, which in our case obtains the form:

$$ln\,[n_1\,(1-n_2)] / [n_2\,(1-n_1)] = -(V(\mathbf{k}_1) / T)\,\eta \qquad (8)$$

In a standard way it can be deduced [10, 11] that equation (8) possesses the non-zero solutions only at temperatures below some critical temperature point $T_{crit}$, determined as

$$T_{crit} = -V(\mathbf{k}_1)\,c\,(1-c). \qquad (9)$$

In the case of continuous disorder-order phase transformation $T_{crit}$ coincides with the disorder-order transition temperature $T_c$. Then from the mentioned above experimental value $T_c[V_2D] = 406$ K it follows that in this compound (for $c = 0.5$)

$$V(\mathbf{k}_1) = -1624\ \text{K}. \qquad (10)$$

Solution of equation (8) with the energy parameter (10) is given in Fig, 2. The obtained $\eta(T)$ dependence indicate that in this system the disorder-order transition is continuous (of the second-order type).

**The model**

We suppose that optic oscillations of hydrogen atoms in the metal lattice in the disordered state (at high temperatures) are determined mainly by the interactions with surrounding metal atoms, while in the ordered state (at low temperatures) the additional H-H interactions become significant and the local potential field around H-atoms modifies that provides corresponding changes of local frequencies.

In order to give the principal picture and some quantitative estimations of the hydrogen-hydrogen interaction influence on the potential field acting on individual hydrogen atoms, we shall apply a well known mathematical scheme of a one-dimensional quantum oscillator vibrating in a potential well described by the function

$$V(x) = Kx^2 / 2. \qquad (11)$$

Corresponding Schrödinger equation has the form

$$-(\hbar^2 / 2m)\,[d^2 / dx^2]\,\psi(x) + (K\,x^2 / 2)\,\psi(x) = \varepsilon\psi(x) \qquad (12)$$

or

$$[d^2 / dx^2]\,\psi(x) + [(2m\,\varepsilon / \hbar^2) - (m\,K\,x^2 / \hbar^2)]\,\psi(x) = 0. \qquad (12')$$

Solution of this equation is well known, and we shall not repeat it here.

The eigenvalues of (12') are

$$\varepsilon \equiv \varepsilon_0 = \hbar\,\omega_0\,(\lambda/2) = \hbar\,\omega_0\,(n + 1/2), \quad \text{where} \quad \omega_0 \equiv \sqrt{(K/m)}. \tag{13}$$

Both experimentally determined high-temperature frequencies (A) are related to the given energy level $\hbar\,\omega_0$, which in the non-cubic (tetragonal) crystalline field splits in two sub-levels:

$$\varepsilon_0^{(i)} = \hbar\,\omega_0^{(i)}\,(n + 1/2), \qquad \omega_0^{(i)} \equiv \sqrt{(K^{(i)}/m)}. \quad i = 1, 2 \tag{14}$$

They can be represented by the corresponding Einstein temperature parameters

$$\theta_E^{(1, 2)} = \varepsilon_0^{(1, 2)} / k_B. \tag{15}$$

In the low-temperature region, where the hydrogen-hydrogen interaction surpasses the thermal energy, and a superstructure is formed, we have to take into account the set of additional H-H forces acting on the vibrating H-atom. <u>We assume that the additional potential $V_{HH}$ which can be ascribed to a given vibrating hydrogen atom is proportional to the $[1 / (Nc)]$ part of the H-H interaction energy $E(\eta; c)$ determined by (7)</u>. Particularly, we assume that

$$V_{HH} = b\,[E(\eta, c) / (Nc)] = b\,[(k_B / c)]\,[V(0)\,c^2 + V(\mathbf{k}_1)\,(\eta\gamma)^2], \tag{16}$$

where $b$ is some temperature-independent numerical factor.

Then equation (12′) will be replaced by the following one

$$[d^2 / dx^2]\,\psi(x) + [(2m\,(\varepsilon - V_{HH}) / \hbar^2) - (m\,K_m\,x^2 / \hbar^2)]\,\psi(x) = 0. \tag{12″}$$

Basing on the numerical values given in the relations (10) and (A) the additional term $V_{HH}$ can not be treated as a perturbation. We shall try to include it in the mass-parameter. The following sequence of relations is obvious:

$$m\,(\varepsilon - V_{HH}) = m\,\varepsilon\,(1 - V_{HH} / \varepsilon) \equiv m^*\varepsilon\,; \qquad m^* \equiv m\,(1 - V_{HH} / \varepsilon).$$

$$m\,K_m \equiv K_m\,[m^* / (1 - V_{HH} / \varepsilon)] = K_m^*\,m^*, \qquad K_m^* = K_m\,(1 - V_{HH} / \varepsilon)^{-1}. \tag{17}$$

Using these notations, equation (12″) will be transformed into an expression quite similar to (12′).:

$$[d^2 / dx^2]\,\psi(x) + [(2m^*\,\varepsilon / \hbar^2) - (m^*\,K_m^*\,x^2 / \hbar^2)]\,\psi(x) = 0, \tag{18}$$

which can be solved in a similar way as (12′) and we shall arrive to an analogous energy spectrum of a perturbed oscillator

$$\varepsilon = \hbar\,\omega_H(T)\,(n + 1/2), \tag{19}$$

where

$$\omega_H(T) = \sqrt{(K_m^* / m^*)} = \sqrt{(K_m / m)}\,[1 - V_{HH} / \varepsilon]^{-1} = \omega_0\,[1 - V_{HH} / \varepsilon]^{-1}, \tag{20}$$

or

$$\omega_H(T) = \omega_0 \left(1 - b\,(k_B/c)\,[V(0)\,c^2 + V(\mathbf{k}_1)\,(\eta(T)\,\gamma)^2]/\varepsilon\right)^{-1}. \tag{20'}$$

On replacing in the latter expressions $\varepsilon$ by $\varepsilon_0 = \hbar\omega_0$ and repeating the above procedure for both split local states, we arrive to the final expressions:

$$\omega_H^{(i)}(T) = \omega_0^{(i)} f_i(T), \qquad i = 1, 2 \tag{21a}$$

$$f_i(T) = \left(1 - b\,[(V(0)\,c + (V(\mathbf{k}_1)/c)\,(\eta(T)\,\gamma)^2)/\theta_E^{(i)}]\right)^{-1}, \tag{21b}$$

where $\omega_0^{(i)}$ and $\theta_E^{(i)}$ are those determined above (see (14) and (15)).

Basing on relations (21) the concentration and temperature variations of hydrogen local mode frequencies in the ordering metal hydrides have to be reproduced by a single uniting expression:

$$\theta_E^{(i)}(T) = \theta_E^{(i)} f_i(T), \qquad i = 1, 2 \tag{22}$$

where the factor $f_i(T)$ is determined by (21b) and $\theta_E^{(i)}$ are the corresponding high-temperature values of the Einstein characteristic temperatures.

### Results

Today we have a lack of unambiguous information concerning the energy constant $V(0)$, and as we are interested mainly in the temperature dependence of local frequencies we shall assume it to be negligible with respect to other energy constants. Within the frames of this rough assumption we can write the following expression for the characteristic temperatures $\theta_E^{(i)}$ of the split local modes

$$\theta_E^{(i)}(T) = \theta_E^{(i)} \left(1 - b\,[(\gamma\,\eta(T))^2 (V(\mathbf{k}_1)/c)/\theta_E^{(i)}]\right)^{-1}. \qquad (i = 1, 2) \tag{23}$$

In order to compare the predicted temperature dependence (23) with the available numerical data for local frequencies of $V_2D$ system [5], we have ascribed to the energy parameter $V(\mathbf{k}_1)$ the estimated value (10) and to the numerical parameter $b$ - three different trial values: $b = 1$, $b = 2$ and $b = 3$.

In Fig. 3 we have presented temperature dependencies of both local frequencies for the three values of the numerical factor $b$. The experimental points shown in the figure are those given in (A) and (B) relations. Calculated curves were normalized on the corresponding experimental points at 425 K

### Concluding remarks

We see that coincidence between the calculated and measured low-temperature $\theta_E^{(i)}$ values is sufficiently good in the case of numerical coefficient $b = 3$. Reduction of both local vibration frequencies in the ordered configuration of hydrogen subsystem has a simple explanation.

From our point of view, the hydrogen-hydrogen forces which provide spatial ordering of the hydrogen subsystem (and become effective at low temperatures) provide formation of additional bonds for each oscillating H-atom that looks like an increase of its inertial mass. As a result, its characteristic frequency within the metal lattice potential wells reduces.

At the same time it must be stressed that any attempts to ascribe some definite physical sense to the best value of the trial numerical coefficient *b* are sufficiently meaningless, as we have neglected: a/ the energy parameter V(0) and b/ the effect of the possible redistribution of hydrogen atoms from tetra-positions to octa-positions on crossing the disorder-order transition point at 406 K.

## Acknowledgements

Investigations were performed within the STCU project N 3867.

**References**

1. N.Stump, G.Alefeld, D.Tocchettti. Optical excitations due to deuterium in niobium. Sol.St.Com.,19, 805-507 (1976).

2. J.J Rush, A.Magerl, J.M.Rowe, J.M.Harris, J.L.Provo. Tritium vibrations in niobium by neutron spectroscopy. Phys. Rev. B, **24**, 4903-4905 (1981)

3. D.Richter, S.M.Shapiro. Study of the temperature dependence of the localized vibrations of H and D in niobium. Phys. Rev. B, **22**, 599-605 (1980)

4. A.Magerl, J.J.Rush, J.M.Rowe. Local modes in dilute metal-hydrogen alloys. Phys. Rev. B, **33**, 2093-2097 (1986)

5. J.M.Rowe. A neutron scattering study of the vibrational and diffusional motions of deuterium in the α and β phases of $VD_{0.5}$. Sol. St. Comm., **11**, 1299-1302 (1972)

6. T.Schober, H.Wenzl. The systems NbH(D), TaH(D), VH(D): Structures, Phase Diagrams, Morphologies, Methods of Preparation. *in* Hydrogen in Metals II, Topics in Applied Physics v. **29**, Springer Verlag, pp. 11-71 (1978).

7. V.A.Somenkov, S.Sh.Shilshtein. Phase Transformations of Hydrogen in Metals. Preprint of I.Kurchatov IAE, Moscow, 1978. (in Russian)

8. Y.Fukai, S.Kazama. NMR studies of anomalous diffusion of hydrogen and phase transition in vanadium-hydrogen alloys. Acta Metall., **25**, 59-69 (1977)

9. T. Schober. Metal-hydrogen phase diagrams. *in* Electronic Structure and Properties of Hydrogen in Metals. (Eds. P.Jena and C.B.Satterthwaite), Plenum Press, N.-Y. and London, 1983..pp. 1 – 10.

10. A.G.Khachaturyan. Theory of Structural Transformations in Solids. Wiley, N.-Y., 1983.

11. I.G. Ratishvili., P.Vajda. Ordering in the system β-$TbH_{2+x}$. Phys. Rev. B **47**, 14062 – 14069 (1993).

12. L.I.Schiff. Quantum mechanics, McGrow-Hill, New-York, Toronto-London, 1955; ch. 4.

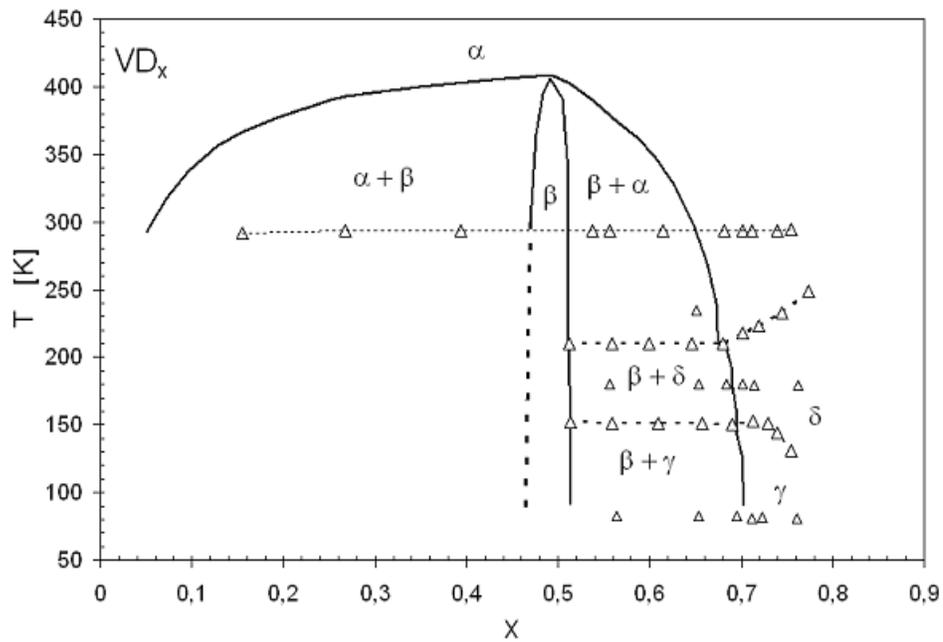

Fig. 1.

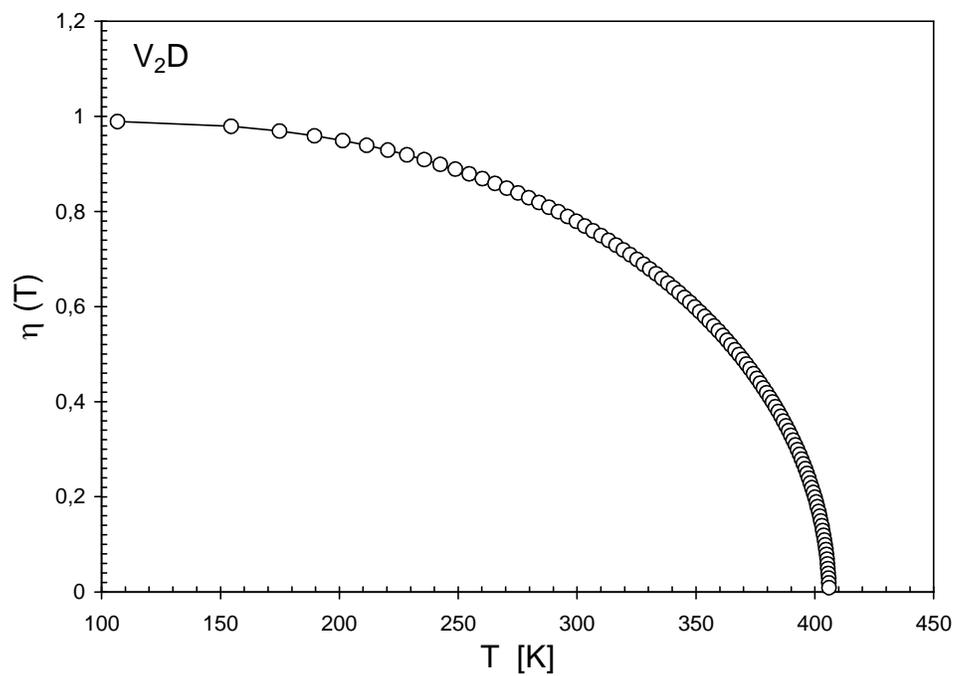

Fig. 2.

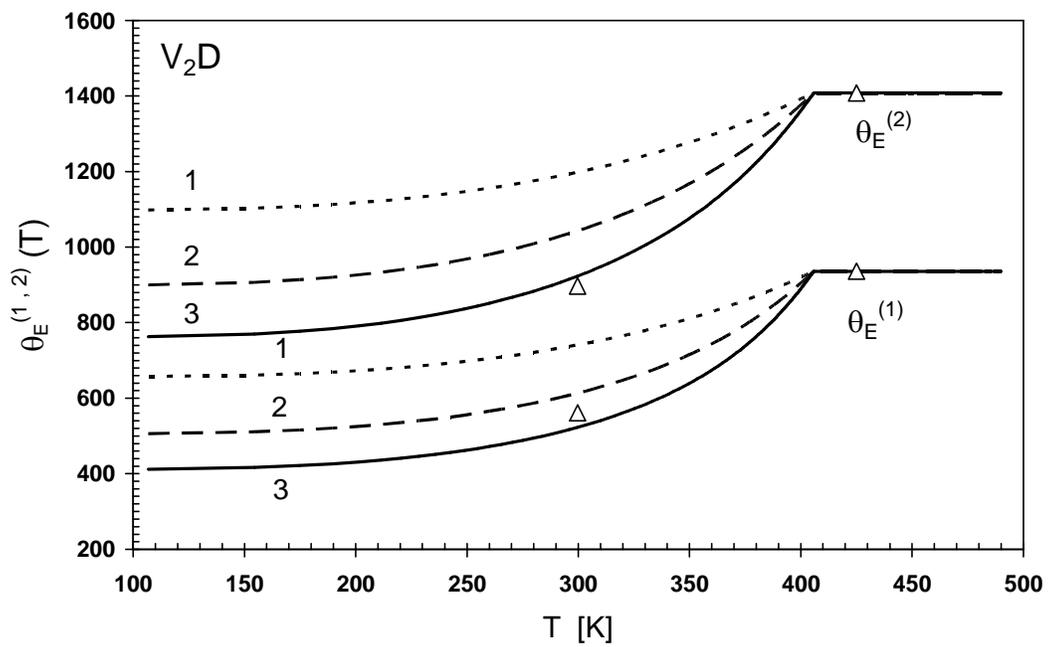

Fig. 3.

**Figure captions**

Fig. 1. Phase diagram of V-D system [9].

Fig. 2. Temperature dependence of the order parameter η. (Calculated η(T) dependence illustrates the continuous character of the disorder-order transition in the hydrogen subsystem).

Fig. 3. Calculated temperature dependence of the local mode frequencies. (Triangles denote experimental points [5]; dashed thin lines, dashed solid lines and thick solid lines correspond, respectively, to the trial numerical coefficient values: $b = 1$, $b = 2$ and $b = 3$).